\documentclass[aps,pra,twocolumn,showpacs,superscriptaddress,10pt]{revtex4-1}

\usepackage{amsmath,amssymb}
\usepackage{bm}
\usepackage{graphicx}
\usepackage{epstopdf}
\usepackage{color}

\usepackage{tikz}
\newcommand*{\circled}[1]{\lower.7ex\hbox{\tikz\draw (0pt, 0pt)%
		circle (.5em) node {\makebox[1em][c]{\small #1}};}}
\usepackage[T1]{fontenc}
\usepackage{tgtermes}
\usepackage[colorlinks ,linkcolor=blue,anchorcolor=blue,citecolor=blue,urlcolor=blue]{hyperref}
\begin{document}
	
	\title{Trion ordering in the attractive three-color Hubbard model on a $\pi$-flux square lattice}
	\author{Xiang Li}
	\thanks{These authors contributed equally to this work}
	\affiliation{School of Physics and Technology, Wuhan University, Wuhan
		430072, China}
	\author{Yumeng Li}
	\thanks{These authors contributed equally to this work}
	\affiliation{School of Physics and Technology, Wuhan University, Wuhan
		430072, China}
	\author{Quan Fu}
	\thanks{These authors contributed equally to this work}
	\affiliation{School of Physics and Technology, Wuhan University, Wuhan
		430072, China}
	\author{Yu Wang}
	\email{yu.wang@whu.edu.cn}
	\affiliation{School of Physics and Technology, Wuhan University, Wuhan
		430072, China}
	
	\begin{abstract}
		Ultracold multicomponent fermions (atoms/molecules) loaded in optical lattices provide an ideal platform for simulating SU($N$) Hubbard models that host unconventional many-body quantum states beyond SU(2). A prime example is the attractive three-color Hubbard model, in which trion states emerge at strong coupling. Nevertheless, much of its trion ordering on two-dimensional lattices remains uncertain. Here, we employ the determinant quantum Monte Carlo (DQMC) method to simulate the attractive three-color Hubbard model on a $\pi$-flux square lattice at half filling. We show that color-dependent attractive interaction can induce coexisting charge density wave (CDW) and N\'eel ordered states in the three-color $\pi$-flux Hubbard model. In particular, enhanced charge fluctuations (cf. honeycomb lattice) cause much stronger N\'eel ordering on the $\pi$-flux square lattice. The coexisting charge and N\'eel orders survive up to a melting temperature, at which they vanish simultaneously. The Ginzburg-Landau (GL) analysis on the coexistence of CDW and N\'eel orders demonstrates how color-dependent Hubbard interactions stabilize coexisting orders from the perspective of GL free energy principle.
	\end{abstract}
	
	\maketitle
	\section{Introduction}

	The Hubbard model offers a prototype Hamiltonian to describe quantum phase transitions in strongly correlated fermionic systems. It simply incorporates the nearest-neighbor hopping and the on-site interaction capturing the essence of the system that leads to a competition between itinerancy and localization, but remains a challenging problem due to the non-perturbative nature of strong correlation physics. The standard SU(2) Hubbard model is commonly used to describe interacting electrons in solids, and interestingly the four-component Hubbard model arises in two-orbital transition-metal oxides~\cite{Pati1998,Corboz2012} and twisted bilayer systems where the layer pseudospin and the real spin are combined together~\cite{Xu2018,Zhang2021}. 
	
	Over the past two decades, fast-evolving ultracold atom technologies have enabled the experimental realization of the multicomponent Hubbard models using large-spin fermions in optical lattices~\cite{jaksch2005cold,lewenstein2007ultracold,Bloch2008,bloch2012quantum,GorshkovSUNtheory,Taie2010,DeSalvo2010,Taiemagcoropt,Ozawamagcoropt,Taie2022,Tusi2022}. Beyond investigations in crystalline solids, large-spin ultracold fermions trapped in optical lattices provide a highly tunable platform for implementing multicomponent Hubbard models, enabling explorations of exotic many-body physics that is rare in solids. So far ultracold atom experiments have realized Mott insulator and antiferromagnetic correlations in the SU(2$N$) Hubbard models ($N$=2, 3) ~\cite{Taiemagcoropt,Ozawamagcoropt,Taie2022}, and color-selective Mott physics in the three-color Hubbard model with tunable inter-color couplings~\cite{Tusi2022}. Theoretical studies of the multicomponent Hubbard models have been boosted by the development of numerical methods and ever-increasing computing power. The quantum Monte Carlo (QMC) method, has emerged as a powerful computational tool for simulating various multicomponent Hubbard models. Specifically, it has been successfully applied to investigate: (1) repulsive SU($N$) Hubbard models ($N$=3, 4, 6)~\cite{Yu2024,Zhou2014,Zhou2016,Zhou2017,Zhou2018,Xu2024} and (2) attractive three-color Hubbard models with both SU(3)-symmetric interaction~\cite{xu2019quantum,li2021,Stepp2025} and tunable inter-color couplings~\cite{li2022}. These theoretical studies focus on two-dimensional geometries, particularly square and honeycomb lattice structures that are experimentally realizable in ultracold atom quantum simulators. In $d=\infty$ dimensions, dynamical mean-field theory (DMFT) has been employed to simulate the attractive three-color Hubbard model with SU(3) symmetry that can be broken via tuning inter-color couplings~\cite{Okanami3SF,Titvinidze2011,Koga2017}.
	
	In the repulsive SU($N$) Hubbard model, large $N$ enhances quantum fluctuations, leading to the appearance of new phases that are not stable or present in the SU(2) case~\cite{WuSUNtheory,Wufluctuation}. But in the attractive SU($N$) Hubbard model, large $N$ suppresses charge fluctuations, making the system classical-like. Nevertheless, unlike the repulsive SU($N$) model and the attractive SU(2) model, multiple spin states in the attractive Hubbard model favor the formation of various fermionic bound states (e.g. trions and doublons)~\cite{Pohlmann2013,Capponi2009,xu2019quantum,li2021,Stepp2025,li2022,Okanami3SF,Titvinidze2011,Koga2017}. The attractive SU(3) Hubbard model, as a minimal SU($N$) model beyond SU(2), hosts both simple bound states and strong charge fluctuations, and therefore offers an opportunity to study unconventional strong correlation physics in which various types of composite particles get involved~\cite{xu2019quantum,li2021,li2022}.
	
	In experiment, the SU(3) $^6$Li Fermi gas with tunable attractive interaction has been realized in a trap~\cite{OttensteinLi,Huckans2009}. Until very recently, the half-filled three-color Hubbard model (with color-dependent Hubbard interactions) was realized with ultracold $^6$Li atoms in a two-dimensional square optical lattice, where the notorious three-body loss that once rendered such system unstable has been dramatically suppressed  \cite{Mongkolkiattichai2025imag}. It has also been proposed that microwave-shielded ultracold dipolar molecules can be used to simulate the attractive SU(3) Hubbard model, since the double microwave shielding technique can suppress loss processes~\cite{Bigagli2024,Dutta2025,karman2025}. Meanwhile, there has been a growing theoretical interest in the ground-state properties of the attractive three-color Hubbard model. In earlier studies, the variational~\cite{Rappvariation,Rappvariation2}, self-energy functional~\cite{InabaSFA,InabaSFA2} and DMFT~\cite{Okanami3SF,Titvinidze2011,Koga2017} methods unanimously suggest that, in $d=\infty$ dimensions a quantum phase transition between a color superfluid phase and a trionic phase can occur in the attractive three-color Hubbard models with both SU(3)-symmetric interaction and tunable inter-color couplings, independent of specific filling. This phase transition is of particular interest, since it is reminiscent of the transition between the quark superfluid and the baryonic phase in high-energy physics~\cite{FodorQCD,AokiQCD,WilczekQCD}. 
	
	Recently, the sign-problem-free QMC simulations of the half-filled attractive three-color Hubbard model show a very different scenario in two dimensions~\cite{xu2019quantum,li2021,li2022}. When the Hubbard interaction is SU(3)-symmetric, the CDW quantum critical point on the honeycomb lattice cannot be described within the Gross-Neveu-Yukawa framework for SU(3) Dirac fermions~\cite{xu2019quantum}, which may be the consequence of forming doublons and trions. Moreover, QMC simulations reveal that the attractive three-color Hubbard model with anisotropic inter-color couplings hosts coexisting CDW and N\'eel phases on the honeycomb lattice, where the N\'eel order is weak and originates from the emergence of magnetic off-site trions~\cite{li2022}. While the coexistence of charge and spin orders occurs in a variety of two-dimensional materials~\cite{Kivelsoncoexistsuper,Lee1997,Tranquada1994,Zhang2020},  the interplay of on-site and off-site trions in the attractive three-color Hubbard model offers a different mechanism for realizing the coexisting CDW and N\'eel orders~\cite{li2022}.
	
	So far the study on the two-dimensional attractive three-color Hubbard model is very limited in the literature due to lack of accurate computational methods. In this work, we employ the DQMC method to simulate the attractive three-color Hubbard model on a $\pi$-flux square lattice at half filling. The $\pi$-flux square lattice features a Dirac-like band structure similar to that of the honeycomb lattice, despite exhibiting an increased coordination number from $z=3$ to $z=4$. We investigate the influence of charge fluctuations on the N\'eel ordering and demonstrate thermal melting of the coexisting CDW and N\'eel orders that is of broad interest in condensed matter physics. We also suggest a GL analysis, which implies that the coexistence of CDW and N\'eel orders is stabilized by color-dependent Hubbard interactions.
	
	The rest of this paper is organized as follows. In Sec.~\ref{sec:model}, we introduce the attractive three-color Hubbard model on a $\pi$-flux lattice and detail DQMC parameters. In Sec.~\ref{sec:coordinate}, we establish that increasing coordination number enhances N\'eel ordering through increased off-site trion density. In Sec.~\ref{sec:bond},  bond-bond correlations show that off-site trions are not long-range correlated and therefore randomly oriented. In Sec.~\ref{finiteT}, we demonstrate simultaneous vanishing of CDW and N\'eel orders at finite temperatures. In Sec.~\ref{sec:GL}, the GL theory reveals that anisotropic interactions stabilize coexisting orders. Conclusions and discussions are presented in Sec.~\ref{sec:conclusion}.
	\section{Model and method}
	\label{sec:model}
	At half filling, the attractive three-color Hubbard model on a $\pi$-flux square lattice is described by the Hamiltonian
	\begin{equation} \label{main.Eq.1}
		\begin{aligned}
			H=&-\sum_{\langle ij\rangle,\alpha}t_{ij}(c^{\dagger}_{i\alpha}c_{j\alpha}+\mathrm{H.c.}) \\
			&+\sum_{i,\alpha<\beta}U_{\alpha\beta}(n_{i\alpha}-\frac{1}{2})(n_{i\beta}-\frac{1}{2}),
		\end{aligned}
	\end{equation}
	where $\langle ij\rangle$ denotes a pair of nearest-neighbor sites; $\alpha, \beta =1, 2, 3$ are the color indices;
	$ c^\dagger_{i\alpha} $ and $ c_{i\alpha} $ are the fermionic creation and annihilation operators for a color-$ \alpha $ state on site $i$; 
	$n_{i\alpha}=c^{\dagger}_{i\alpha}c_{i\alpha}$ is the particle number operator; 
	$U_{\alpha \beta} < 0 $ is the on-site inter-color attraction. The chemical potential vanishes at half filling. 
	We introduce the color-dependent Hubbard interactions, setting $U_{12}=U$ and $U_{13}=U_{23}=U^{\prime}$. 
		\begin{figure}[t]
		\centering
		\includegraphics[width=0.99\linewidth]{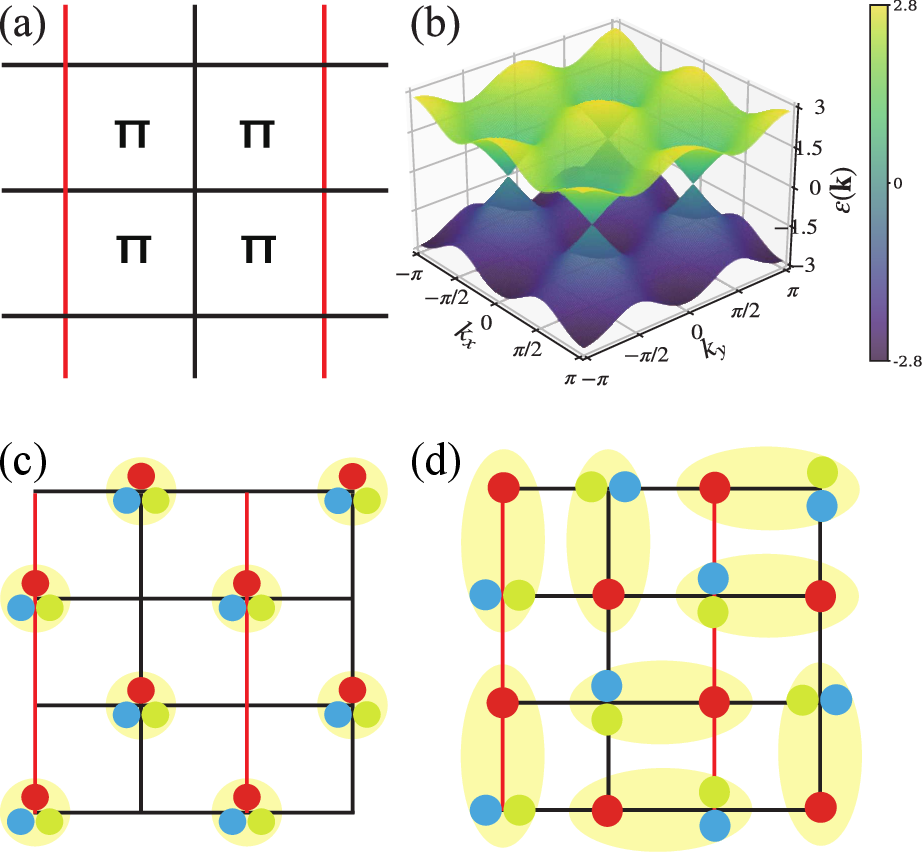}
		\caption{
			(a) Schematic illustration of the $\pi$-flux square lattice. Black and red lines denote hopping integrals $t$ and $-t$, respectively, resulting in a $\pi$ flux per plaquette.
			(b) Band structure of the $\pi$-flux square lattice with Dirac points at momentum-space locations $(k_x,k_y)=(\pm\frac{\pi}{2},\pm\frac{\pi}{2})$.
            (c) Schematic illustration of the CDW order and on-site trions. (d) Schematic illustration of the N\'eel order and off-site trions. 
		}
		\label{lattice}
	\end{figure}
	
	On the $\pi$-flux square lattice, the gauge ($t_{\hat{e}_{x}}=t$, $t_{\hat{e}_{y}}=(-1)^{x} t$) is chosen for the nearest-neighbor hopping integral $t_{ij}$, such that the product of phases of hopping integrals around a plaquette equals $e^{i\pi}=-1$, as illustrated in Fig.~\ref{lattice}(a). The dispersion relation of the $ \pi $-flux square lattice is given by $ \epsilon(\mathbf{k}) = \pm 2t\sqrt{\cos^2k_x + \cos^2k_y} $, and the band structure is illustrated in Fig.~\ref{lattice}(b). At half filling, the Fermi surface touches four Dirac points. This gives rise to a Dirac-like band structure similar to that of the honeycomb lattice at half filling, except that the coordination number increases from three to four. This makes the $\pi$-flux square lattice an ideal platform for investigating how the coordination number alone influences the coexistence of the CDW order developed mainly by on-site trions (schematically shown in Fig.~\ref{lattice}(c)) and the N\'eel order developed by off-site trions (a local bound state with a one-fermion end on one site and a two-fermion end on the nearest-neighbor site, schematically shown in Fig.~\ref{lattice}(d)) \cite{li2022}.
	
	At half filling, the DQMC simulation of the attractive three-color Hubbard model defined by Eq.~(\ref{main.Eq.1}) is sign-problem-free when the Hubbard-Stratonovich decomposition is employed in the color-flip channel ~\cite{Wang2015,xu2019quantum,li2021,li2022}. Our simulations are performed on the $\pi$-flux square lattices with sizes $L=6,8,10,12,14,16$. Unless specifically stated, the temperature $T$ and Hubbard $U_{\alpha\beta}$ are given in the unit of hopping integral $t$. In our simulations, we set $|U|=6$ throughout this work. The temperature is set to $T=0.1$ in Secs.~\ref{sec:coordinate} and \ref{sec:bond}, while finite-temperature properties in Sec.~\ref{finiteT} are investigated across various $T$.
	
	\section{off-site trion formation and N\'eel ordering
     }
	\label{sec:coordinate}
	The QMC simulations of attractive three-color fermions on the honeycomb lattice have demonstrated that charge fluctuations have a significant influence on the density of off-site trions \cite{xu2019quantum,li2021,li2022}. In this section, we will increase coordination number from 3 (honeycomb lattice) to 4 ($\pi$-flux square lattice), and investigate the corresponding influence on the N\'eel order developed by the off-site trions.
	
	Off-site trions are transformed from on-site trions as a consequence of charge fluctuations and their density scales as $t^2/|U|^2$~\cite{li2022}. To characterize the on-site trion structure, we define the on-site triple occupancy
	\begin{equation}
		P_{3}=\frac{1}{N}\sum_{i}\langle n_{i1}n_{i2}n_{i3}\rangle,\label{eq:P3}
	\end{equation} 
	where $N=L^2$ is the number of lattice sites. The structure of off-site trions can be described by the off-site triple occupancies
	\begin{equation}
		\begin{aligned}
			P_{3\mathrm{off};1}&=\frac{1}{4N}\sum_{\langle ij\rangle}\langle n_{i2}n_{i3}n_{j1}\rangle,\\
			P_{3\mathrm{off};3}&=\frac{1}{4N}\sum_{\langle ij\rangle}\langle n_{i1}n_{i2}n_{j3}\rangle.
		\end{aligned}
	\end{equation}
	\begin{figure}[t]
		\centering
		\includegraphics[width=0.85\linewidth]{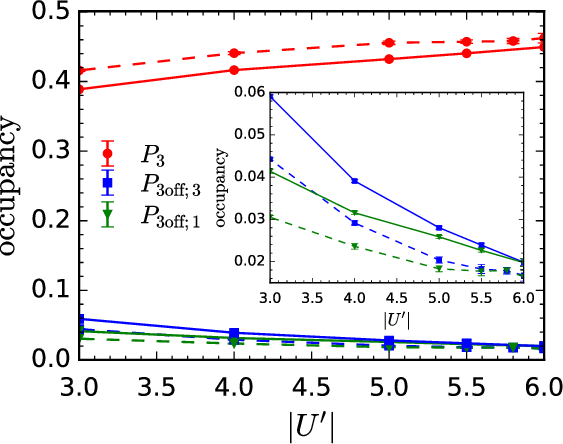}
		\caption{The on-site triple occupancy $P_3$ and off-site triple occupancies $P_{3\mathrm{off};1}$, $P_{3\mathrm{off};3}$ are plotted as functions of $|U^{\prime}|$. Solid and dashed curves correspond to the data on the $L=12$ $\pi$-flux square lattice and $L=9$ honeycomb lattice (extracted from Ref.~\cite{li2022}), respectively. A zoom-in view of $P_{3\mathrm{off};1}$ and $P_{3\mathrm{off};3}$ curves is shown in the inset.}
		\label{occupation}
	\end{figure}
	The triple occupancies $P_{3}$, $P_{\mathrm{3off};1}$, and $P_{\mathrm{3off};3}$ as functions of $|U^{\prime}|$ on the $\pi$-flux square lattice are plotted as solid curves in Fig.~\ref{occupation}. As a comparison, we also extract the  $P_{3}-|U^{\prime}|$, $P_{\mathrm{3off};1}-|U^{\prime}|$, and $P_{\mathrm{3off};3}-|U^{\prime}|$ data for the honeycomb lattice from Ref.~\cite{li2022}, and plot them as dashed curves in Fig.~\ref{occupation}. Crucially, at a fixed $|U^{\prime}|$, $P_{3}$ on the $\pi-$flux square lattice is significantly lower than that on the hoenycomb lattice, while $P_{3\mathrm{off}}$ on the $\pi-$flux square lattice is significantly higher than that on the honeycomb lattice. 
	This demonstrates that increasing the coordination number by 1 enhances charge fluctuations, promoting off-site trion formation from on-site trions. Another key observation from Fig.~\ref{occupation} is that the difference between $P_{3\mathrm{off};1}$ and $P_{3\mathrm{off};3}$ at a fixed $|U^{\prime}|$ on the $\pi$-flux square lattice is larger than that on the honeycomb lattice, which manifests that both ends of the off-site trion carry larger net magnetic moments with increasing coordination number.
	\begin{figure}[t]
		\includegraphics[width=0.99\linewidth]{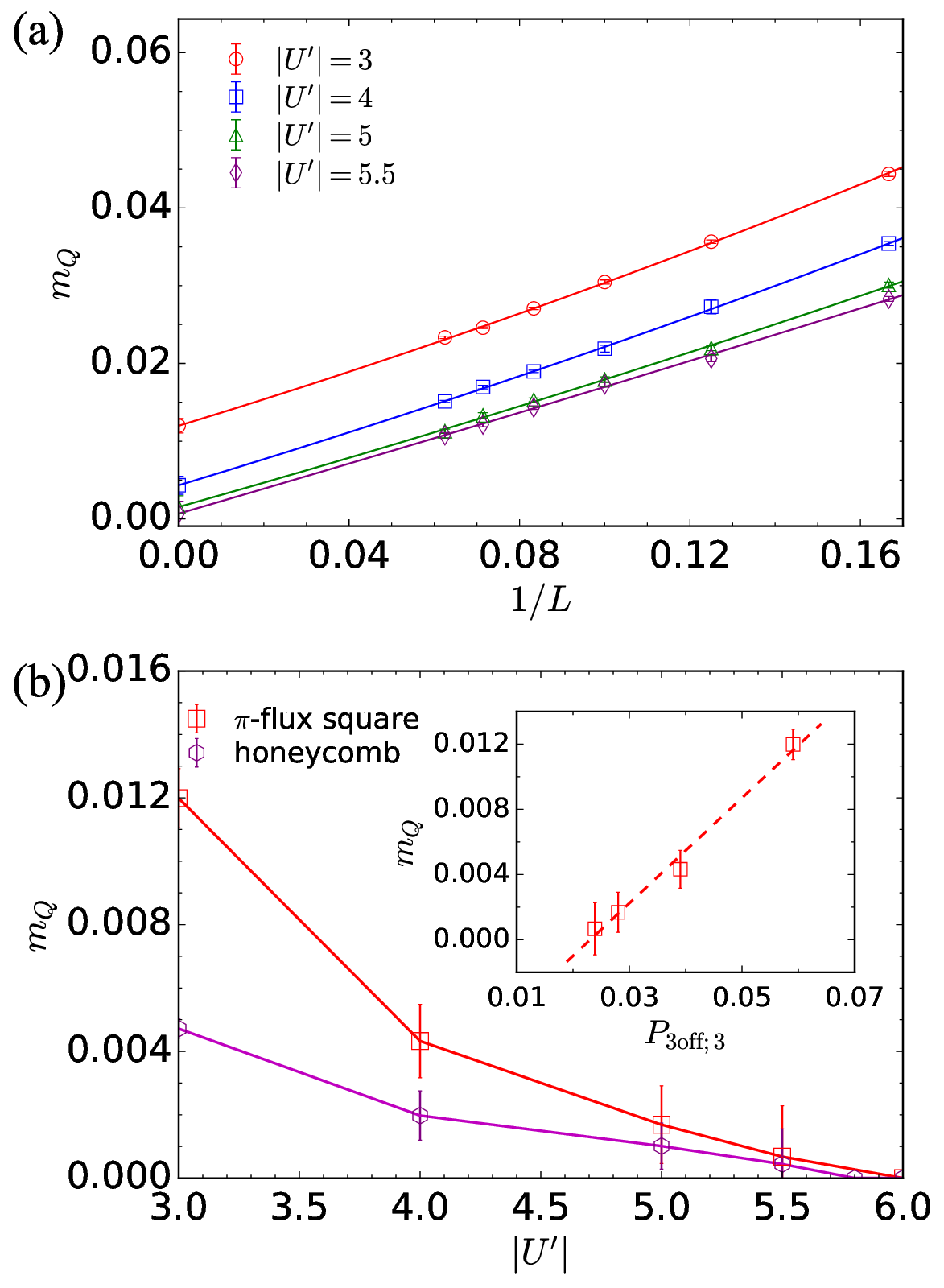}
		\caption{
			(a) The finite-size extrapolations of the N\'eel order parameter $m_Q$ for various $|U^{\prime}|$. The quadratic polynomial fitting is used.
			(b) The N\'eel order parameter $m_Q$ is plotted as a function of $|U^{\prime}|$ on the $\pi$-flux square lattice (red squares) and honeycomb lattice (extracted from Ref.~\cite{li2022}) (purple hexagons). The inset shows the nearly linear relationship between $m_Q$ and $P_{3\text{off};3}$ on the $\pi$-flux square lattice, indicating enhanced N\'eel order with increased off-site trion density.
		}
		\label{extrapolation orders}
	\end{figure}
	
	We then investigate the impact of coordination number on the strength of N\'eel order. According to Refs.~\cite{Wang2014,li2022,Scalettar2023su3r}, the N\'eel order parameter is defined as
	\begin{equation}
		m_{Q}=\frac{1}{N}\sqrt{\sum_{ij}(-1)^{i_x+i_y+j_x+j_y}\langle m_{i}m_{j}\rangle}.\label{eq:neel}
		\end{equation}
	Here $(i_x,i_y)$ is the coordinate of site $i$; $m_{i}=\frac{1}{4}\left(n_{i1}+n_{i2}-2n_{i3}\right)$, which defines the magnetic moment at site $i$. $m_{i}$ vanishes when site $i$ contains three colors (123), while it is non-zero when site $i$ accommodates either net colors (12) or net color 3. We perform finite-size extrapolations of the N\'eel order parameter $m_{Q}$ at $|U^{\prime}|=3,4,5,5.5$, as presented in Fig.~\ref{extrapolation orders}(a). The N\'eel order develops when $|U^{\prime}|$ is below 5.5. Based on the finite-size extrapolations, we obtain the thermodynamic-limit values of $ m_{Q} $ and plot the $m_{Q}-|U^{\prime}|$ relation on the $\pi-$flux square lattice in Fig.~\ref{extrapolation orders}(b). As a comparison, we also extract the $m_{Q}-|U^{\prime}|$ data on the honeycomb lattice from Ref.~\cite{li2022} and plot this dataset in Fig.~\ref{extrapolation orders}(b). The inset of Fig.~\ref{extrapolation orders}(b) reveals a nearly linear relationship between $m_Q$ and $P_{3\mathrm{off};3}$, demonstrating that enhanced N\'eel ordering results from the increased off-site trion density. For a given $|U^{\prime}|(<6) $,  increasing coordination number from 3 (honeycomb lattice) to 4 ($\pi$-flux square lattice) induces much stronger N\'eel order. Fig.~\ref{extrapolation orders}(b) also illustrates the vanishing of the N\'eel order $(m_{Q})$ as $|U^{\prime}| \to 6 $, accompanied by the restoration of SU(3) symmetry in the three-color Hubbard model.


\section{Spatial correlations of off-site trions}\label{sec:bond}
	
	Color-dependent Hubbard interactions can induce coexisting on-site trions and N\'eel-like off-site trions on a half-filled lattice~\cite{li2022}. On the honeycomb lattice, off-site trions have no specific spatial orientation. However, it is unclear whether the $\pi$-flux gauge field would enable off-site trions to acquire a preferential spatial orientation, since the  N\'eel order defined by Eq.~(\ref{eq:neel}) cannot distinguish between two scenarios: (1) off-site trions orient randomly along all four lattice directions; (2) off-site trions exhibit preferential alignment along the $x$-direction compared to the $y$-direction, owing to the uniform hopping integral $t$ along the $x$-direction, in contrast to the alternating $t$ and $-t$ along the $y$-direction. Nevertheless, the difference between these two scenarios can be distinguished by measuring bond-bond correlations. We define the nearest-neighbor kinetic bond operator $ d_{i,\hat{e}_{a}} $ for one off-site trion as
	\begin{equation}
		d_{i,\hat{e}_{a}}=\sum_{\alpha=1}^{3}t_{i,i+\hat{e}_{a}}(c^{\dagger}_{i,\alpha}c_{i+\hat{e}_{a},\alpha}+\mathrm{H.c.}),
	\end{equation}
	where $\hat{e}_{a}$ denotes the two possible nearest-neighbor bond directions, taking the vector forms $\hat{e}_{x}=(1,0)$ or $\hat{e}_{y}=(0,1)$. It should be noticed that on-site trions make small contributions to $d_{i,\hat{e}_{a}}$ since they must be dissociated in order to contribute. The correlation between the bond along the $\hat{e}_{a}$ direction and the bond along the $\hat{e}_{b}$ direction is defined as
	\begin{equation}
		B_{ab}(i,j)=\langle d_{i,\hat{e}_{a}}d_{j,\hat{e}_{b}}\rangle-\langle d_{i,\hat{e}_{a}}\rangle\langle d_{j,\hat{e}_{b}}\rangle.
	\end{equation}
		\begin{figure}[t]
		\centering
		\includegraphics[width=1\linewidth]{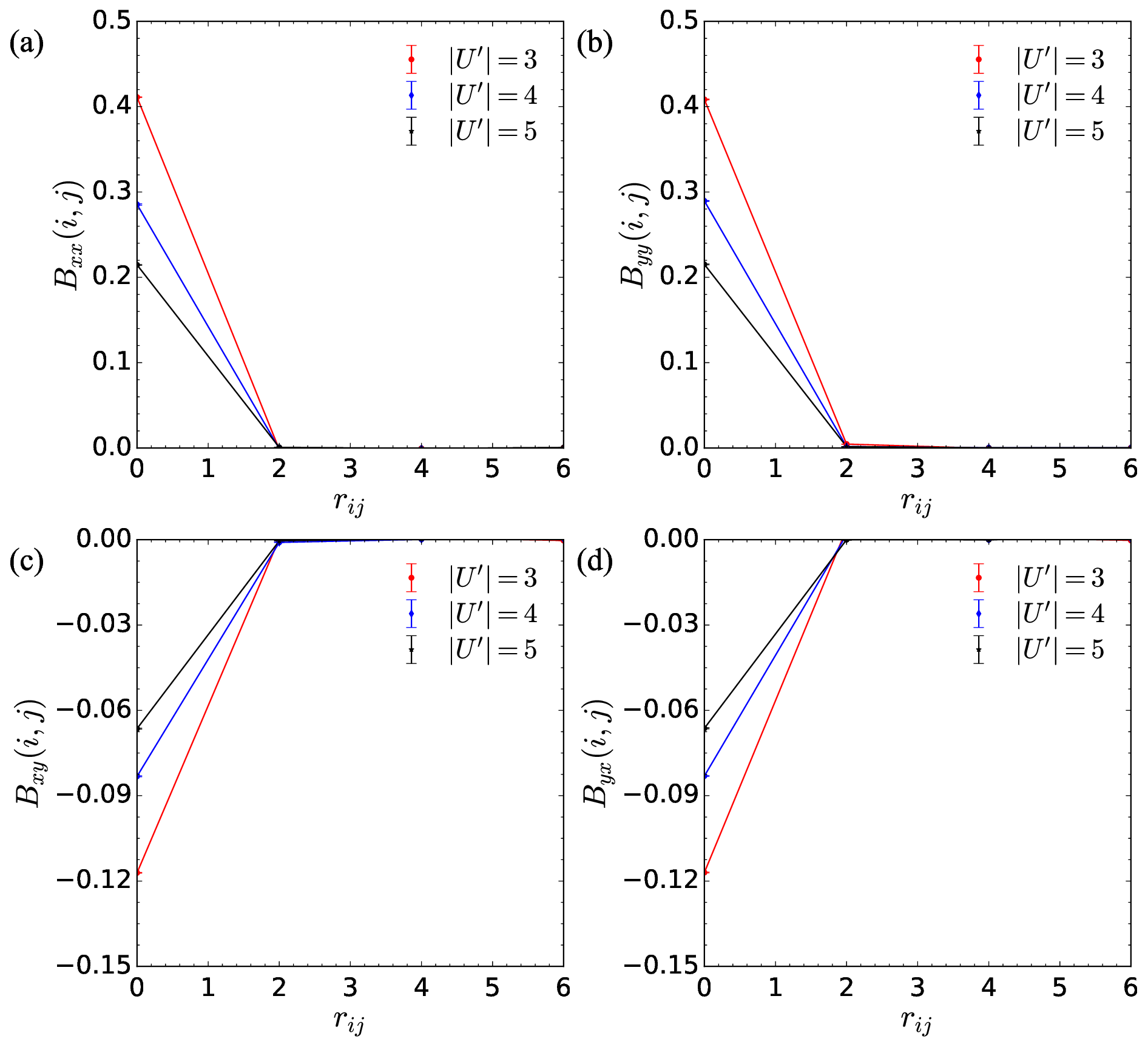}
		\caption{
			The bond-bond correlations (a) $B_{xx}(i,j)$, (b) $B_{yy}(i,j)$, (c) $B_{xy}(i,j)$, and (d) $B_{yx}(i,j)$ are plotted as functions of the distance $r_{ij}$ between sites $i$ and $j$ for various $|U^{\prime}|$ on a $L=14$ lattice.
		}
		\label{bondcor}
	\end{figure}
	In Fig.~\ref{bondcor} we plot the bond-bond correlations $B_{ab}(i,j)$ as functions of $r_{ij}$ for various $|U^{\prime}|$ with $L=14$. The rapid decay of $B_{ab}(i,j)$ to zero demonstrates that off-site trions lack long-range spatial correlations in any direction. The absence of numerical differences between $B_{xx}(i,j)$ and $B_{yy}(i,j)$ shown in Fig.~\ref{bondcor} confirms that off-site trions exhibit no specific spatial orientation even in short range. The above results verify that off-site trions distribute randomly along all four directions. In addition to this primary feature, Fig.~\ref{bondcor} reveals that at a fixed small $r_{ij}$, $B_{ab}(i,j)$ increases with larger interaction anisotropy (smaller $|U^{\prime}|$) , which reflects the enhanced density of off-site trions in stronger anisotropic regimes. The negative values of $B_{xy}(i,j)$ and $B_{yx}(i,j)$ originate from the gauge choice where $t_{\hat{e}_{x}}=t$ and $t_{\hat{e}_{y}}=(-1)^{x} t$. The calculations in this section demonstrate that the N\'eel order is spatial modulation of off-site trions without any preferential orientation.
	
	\section{CDW and N\'eel orders at finite temperatures}
	\label{finiteT}

	This section investigates the thermal melting of coexisting CDW and N\'eel orders through calculating order parameters and correlation functions. The charge order is characterized by the CDW order parameter $D$, which captures charge density modulation, and the staggered order parameter $M_{\alpha}$, quantifying spatial modulation of each fermionic color $\alpha$. Following Ref.~\cite{li2022}, order parameters $D$ and $M_{\alpha}$ are defined as
	\begin{equation}
		\begin{aligned}
			D & = \frac{1}{3N} \sqrt{ \sum_{i,j} (-1)^{i_x+i_y+j_x+j_y} \langle C(i,j) \rangle }, \\
			M_{\alpha} & = \frac{1}{N} \sqrt{ \sum_{i,j} (-1)^{i_x+i_y+j_x+j_y} \langle n_{i\alpha} n_{j\alpha} \rangle },
		\end{aligned}
		\label{eq:order_params}
	\end{equation}
	where $ C(i,j) = \sum_{\alpha,\beta} \langle n_{i\alpha} n_{j\beta} \rangle $ is the density-density correlation function.
	\begin{figure}[t]
		\centering
		\includegraphics[width=\linewidth]{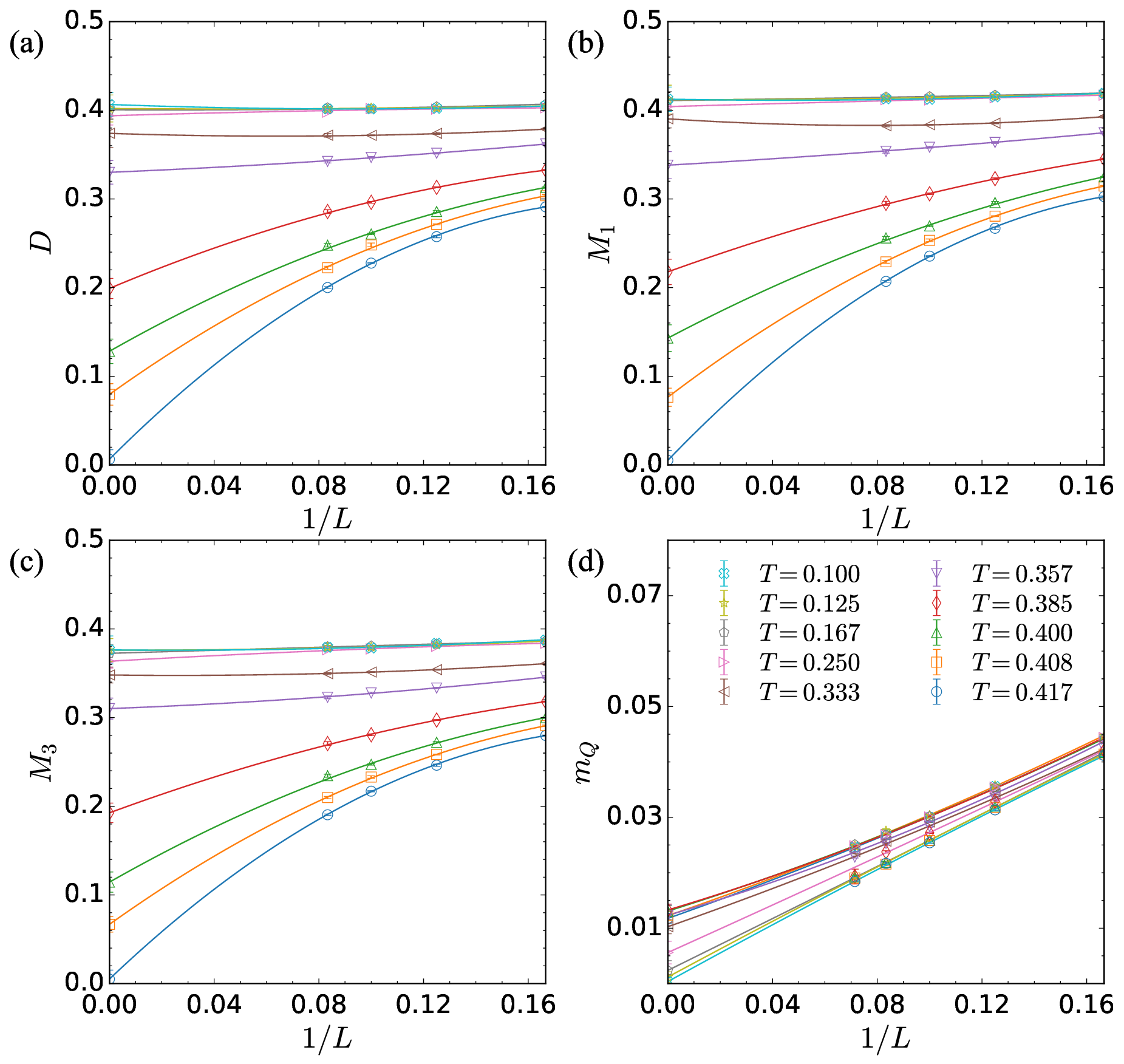}
		\caption{
			The finite-size extrapolations of order parameters (a) $D$, (b) $M_1$, (c) $M_3$, and (d) $m_Q$ at $|U^{\prime}| = 3$ for various temperatures $T$. 
			Order parameters $D$, $M_1$, and $M_3$ characterize charge spatial modulation, while $m_Q$ quantifies N\'eel ordering. The quadratic polynomial fitting is used.
		}
		\label{fss}
	\end{figure}
	
\begin{figure}[t]
		\centering
		\includegraphics[width=1\linewidth]{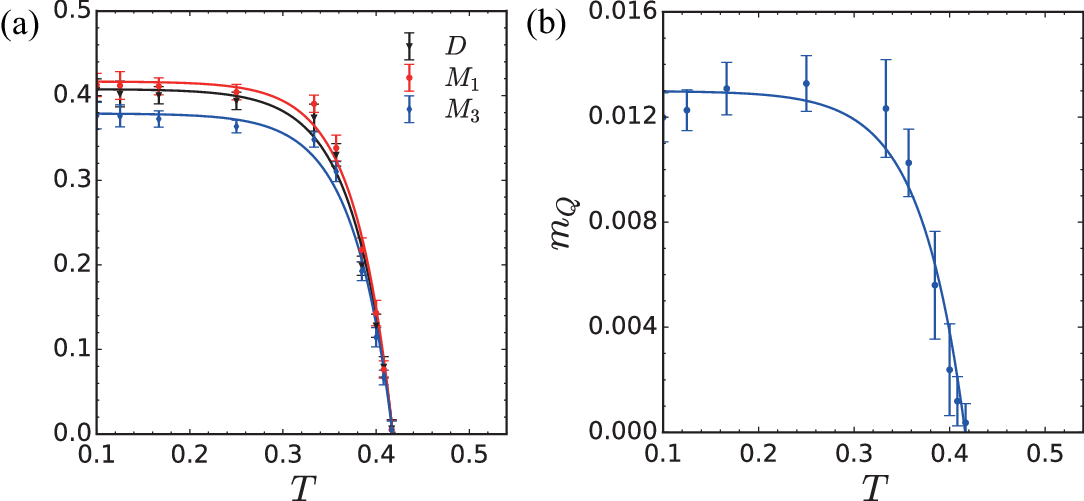}
	\caption{
		The temperature evolution of order parameters at $|U'|=3$: 
		(a) CDW order parameter $D$ and staggered order parameters $M_1$, $M_3$, 
		(b) N\'eel order parameter $m_Q$.
	}
		\label{orders}
	\end{figure}

	We perform finite-size extrapolations of the order parameters $D$, $M_1$, $M_3$, and $m_Q$ at $|U^{\prime}|=3$ across various temperatures $T$, as presented in Fig.~\ref{fss}. Through finite-size extrapolations, we obtain the thermodynamic-limit values of $D$, $M_1$, $M_3$, and $m_Q$ and plot these order parameters as functions of temperature $T$ in Fig.~\ref{orders}. The thermal evolution curves demonstrate the simultaneous vanishing of CDW and N\'eel orders at the critical temperature $T_c = 0.417$, which is within the reach of current cold-atom experiment with attainable low temperature $T\sim 0.25$ \cite{mazurenko2017T}. This simultaneous vanishing originates from the microscopic mechanism that N\'eel order is induced by charge fluctuations of CDW background. Below $ T_c $, the reduced numerical difference between $ M_1 $ and $ M_3 $ with increasing $T$ reflects a decline in off-site trion density. The magnitude of $ D $ remains consistently intermediate between $ M_1 $ and $ M_3 $ since it averages over all colors. 
	
\begin{figure}[t]
		\centering
		\includegraphics[width=0.865\linewidth]{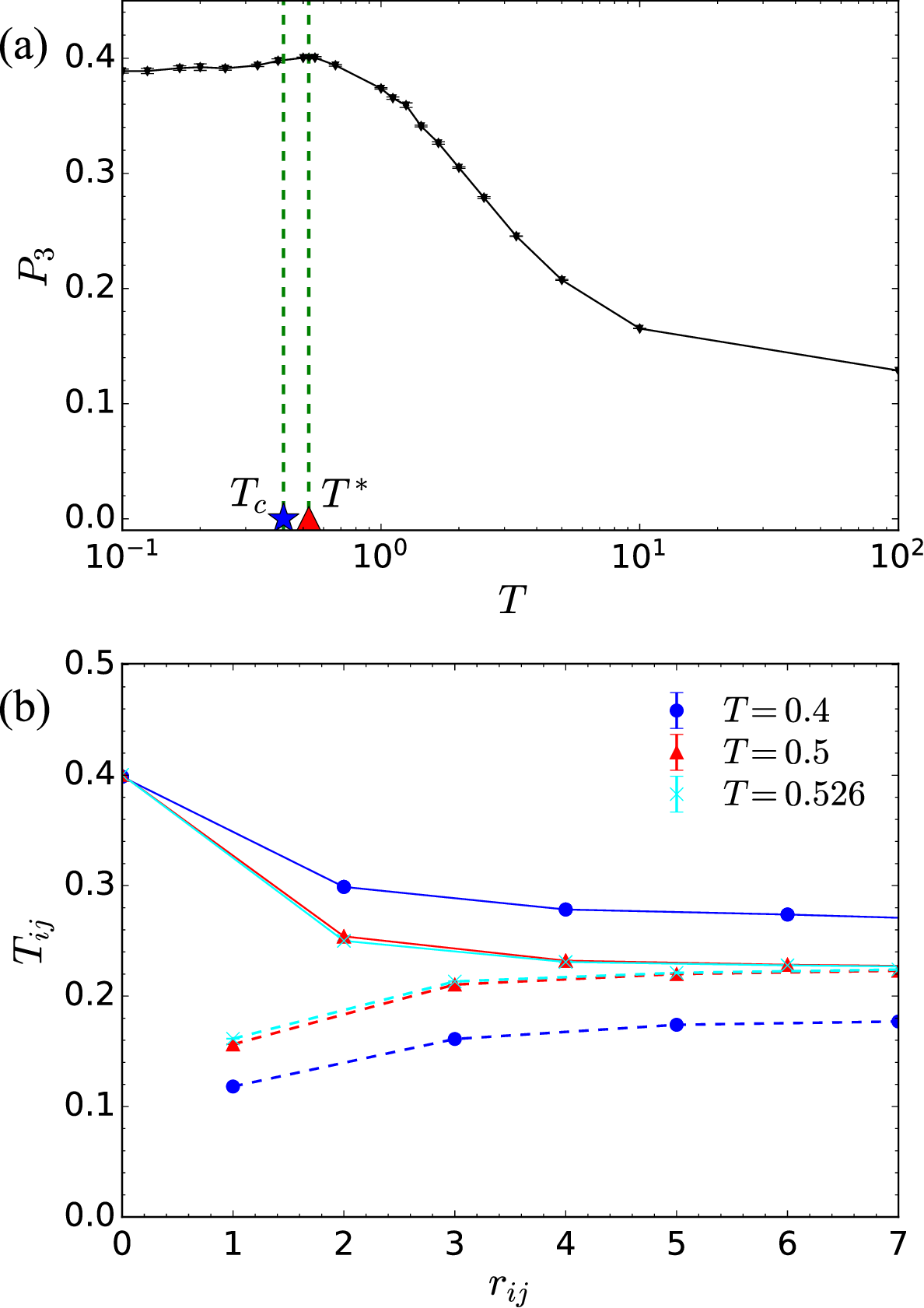}
		\caption{
			(a) The on-site triple occupancy $P_3$ as a function of temperature $T$ at $|U^{\prime}|=3$, with critical temperature $T_c$ (blue pentagram) and peak temperature $T^*$ (red triangle) explicitly marked. 
			(b) Curves of the correlation function $T_{ij}$ are plotted as functions of distance $r_{ij}$ at $|U^{\prime}|=3$ for various temperatures $T$. Solid and dashed curves correspond to $T_{0,2m}$ and $T_{0,2m+1}$, respectively. 
			The lattice size is $L=12$ for both panels.
		}
		\label{p3Tij}
	\end{figure}

	
	To study the thermal evolution and the interplay between on-site and off-site trions, we perform simulations on the temperature dependence of on-site triple occupancy $P_3$ (defined in Eq.~\eqref{eq:P3}). The $P_3$-$T$ curve is plotted in Fig.~\ref{p3Tij}(a), in which $P_3$ exhibits non-monotonic behavior, achieving a maximum at the peak temperature $T^*$ slightly above the critical temperature $T_c$. For $T > T^*$, $P_3$ decreases monotonically with increasing temperature, ultimately converging to the high-temperature limit of 0.125.
	
	Below the critical temperature $T_c$, the triple occupancy $P_3$ increases monotonically with $T$ since the off-site trion density decreases. Remarkably, $P_3$ remains constrained below the theoretical maximum value 0.5 due to charge fluctuations: within the trionic CDW phase, fermions undergo hoppings from on-site trions to adjacent lattice sites, a process that transforms local triple occupancies into off-site trion configurations. Between $T_c$ and $T^*$, $P_3$ continues to increase, due to the melting process from the trionic CDW phase to the trion liquid phase. During this phase transition, on-site trions become randomly distributed throughout the lattice, reducing available hopping channels. The hopping channel reduction suppresses the formation of off-site trions, thereby preserving more on-site trion configurations. With further temperature increase ($T > T^*$), on-site trions dissociate into unbound fermions~\cite{li2021}, resulting in a decrease of $P_3$ toward the high-temperature limit of 0.125.
	
	To further investigate the trion formation during thermal melting, we calculate the correlation function
	\begin{equation}
		T_{ij} = \langle n_{i1}n_{i2}n_{j3} \rangle,
		\label{eq:tij_corr}
	\end{equation}
	which quantifies spatial correlations between the color-3 fermion at site $j$ and the color-(1,2) fermions at site $i$. Since the thermal melting of CDW/N\'eel orders accompanies the restoration of the discrete translational symmetry along the y direction (the change of spatial periodicity $2a\rightarrow a$, with $a$ being the lattice constant), the trion distribution along the y direction should exhibit a corresponding periodicity modulation. Specifically, site $i$ is set as the origin $(0,0)$, and site  $j$ is chosen along the y direction. $T_{ij}$ is then designated as $T_{0,2m}$ and $T_{0,2m+1}$ respectively for site $j=(0, 2m)$ and site  $j=(0, 2m+1)$. As plotted in Fig.~\ref{p3Tij}(b), the correlation function $T_{ij}$ exhibits distinct spatial correlation behaviors distinguishing the ordered trionic density modulation and the disordered phase. At low temperature ($T = 0.4 < T_c$), in the coexisting CDW and N\'eel ordered phase, the inequality $T_{0,2m} > T_{0,2m+1}$ at large $r_{ij}$ signifies ordered spatial modulation of on-site trions, directly manifesting the CDW background. As temperature increases beyond $T_c$ to $T = 0.526$, the equivalence $T_{0,2m} = T_{0,2m+1} = 0.225$ confirms random distribution of on-site trions, signaling the emergence of a trion liquid phase. The deviation from the theoretical asymptotic value 0.25 demonstrates dominant yet non-exclusive on-site trion configurations that coexist with off-site trions and doublons, originating from persistent quantum fluctuations. 
	Collectively, the temperature evolution of on-site triple occupancy $P_3$ and  correlation function $T_{ij}$ provides microscopic signatures of transformation in trionic configurations across phase transitions.

	\section{Ginzburg-Landau theory}
	\label{sec:GL}
	
	Numerical and microscopic analysis in Sec.~\ref{finiteT} reveals simultaneous vanishing of coexisting orders since the N\'eel order arises from charge fluctuations of the CDW order. In this section, we employ a GL framework~\cite{Shou-ChengZhang1997,Nayak2000,Demler2004,Vorontsov2010,Fernandes2010,coleman2015introduction} and identify parameters that stabilize the coexistence of CDW and N\'eel orders, providing a macroscopic and energetic view of the system. 
	
	Without loss of generality, we take $ U_{12}=U$ and $ U_{13}=U_{23}=U^{\prime}$, with the dimensionless parameter $s=|U^{\prime}|/3|U|$ characterizing the interaction anisotropy. 
	By defining operators $\hat{m}_i = n_{i1} + n_{i2} - 2n_{i3}$ and $ \hat{d}_i = n_{i1} + n_{i2} + n_{i3}$, the Hamiltonian defined by Eq.~\eqref{main.Eq.1} is reformulated as
	\begin{equation}
		\begin{aligned}
			\hat{H} = & -\sum_{\langle ij\rangle,\alpha}(\tilde{t}_{ij}\hat{c}^{\dagger}_{i\alpha}\hat{c}_{j\alpha} + \mathrm{H.c.}) \\
			& + \sum_{i} \left[ \frac{1}{2}\left(\gamma - s + \frac{1}{3}\right) \hat{m}_{i}^{2} \right. \\
			& + \left. \frac{1}{2}\left( \gamma + s + \frac{2}{3}\right) \hat{d}_{i}^{2} - \gamma \hat{m}_{i}\hat{d}_{i} \right],
		\end{aligned}
		\label{eq:GL_hamiltonian}
	\end{equation}
	where $\tilde{t}_{ij} = {t}_{ij}/|U|$ and $\gamma$ is a tunable dimensionless parameter. The conditions $\gamma - s + \frac{1}{3} < 0$ and $\gamma + s + \frac{2}{3} < 0$ are imposed to ensure attractive interactions.
	
	The partition function is expressed in a path integral as $Z = \int \mathcal{D}[\bar{c}, c]  e^{-S[\bar{c}, c]}$ with action $S[\bar{c}, c] = \int d\tau  \left( \bar{c}  (\partial_\tau + h) c + \mathcal{H}_{U}(\bar{c}, c) \right)$, where $h$ describes non-interacting fermions on the $\pi$-flux square lattice. The Hubbard-Stratonovich transformation introduces auxiliary fields $m_i$ and $d_i$ to decouple interactions, giving $Z = \int \mathcal{D}[\bar{c}, c, m, d]  e^{-S[\bar{c}, c, m, d]}$. The transformed action then reads
\begin{equation}
	\begin{aligned}
		S[\bar{c}, c, m, d] = & \int_0^{\beta} d\tau \sum_{\mathbf{k},\alpha} \bar{c}_{\mathbf{k},\alpha} (\partial_\tau + \epsilon_{\mathbf{k}}) c_{\mathbf{k},\alpha} \\
		& + \sum_{i} \left( \frac{m_{i}^2}{2(\gamma-s+\frac{1}{3})} + \frac{d_{i}^2}{2(\gamma+s+\frac{2}{3})} \right. \\
		& \quad \left. - a{m}_{i}\hat{m}_{i} - b{d}_{i}\hat{d}_{i} \right).
	\end{aligned}
	\label{eq:transformed_action}
\end{equation}
	Coefficients $a$ and $b$ originate from the mean-field decomposition of the term $\hat{m}_{i}\hat{d}_{i}$ and can be determined via saddle-point equations. At this point, we can integrate out the fermionic fields, yielding the effective partition function $Z = \int \mathcal{D}[m, d]  e^{-S_{E}[m, d]}$ with $ S_{E}[m, d] $ denoting the effective action. The exponential weight is then given by
	\begin{equation}
		\begin{aligned}
			e^{-S_{E}[m, d]} = & \int \mathcal{D}[\bar{c}, c]  e^{-S[\bar{c}, c, m, d]} \\
			= & \det[\partial_{\tau} + h_E] \exp \left[ -\sum_{i} \int_{0}^{\beta} d\tau \right. \\
			& \left. \left( \frac{m_{i}^2}{2(\gamma-s+\frac{1}{3})} + \frac{ d_{i}^2}{2(\gamma+s+\frac{2}{3})} \right) \right],
		\end{aligned}
		\label{eq:eff_SE}
	\end{equation}
	where we introduce the effective Hamiltonian
	\begin{equation}
		\begin{aligned}
			h_{E} = -\sum_{\langle ij \rangle, \alpha} \tilde{t}_{ij}(c_{i,\alpha}^{\dagger} c_{j,\alpha} + \text{H.c.}) 
			 - \sum_{i} \left( a {m}_{i}\hat{m}_{i} + b {d}_{i}\hat{d}_{i} \right).
		\end{aligned}
		\label{eq:eff_hamiltonian}
	\end{equation}
	Saddle-point equations are derived by extremizing the effective action $ S_{E}[m, d] $
	\begin{align}
		\frac{\partial S_E}{\partial m_i} & = \frac{m_i}{\gamma-s+\frac{1}{3}} - a \langle \hat{m}_i \rangle = 0, \label{eq:saddle_m} \\
		\frac{\partial S_E}{\partial d_i} & = \frac{d_i}{\gamma+s+\frac{2}{3}} - b \langle \hat{d}_i \rangle = 0. \label{eq:saddle_d}
	\end{align}
	To establish correspondence between auxiliary fields $(m_i, d_i)$ and physical order parameters $\langle \hat{m}_i \rangle$, $\langle \hat{d}_i \rangle$, we set the coefficients as $a = b/\sqrt{3} = \sqrt{3}$. Fourier transforming the magnetic term $\sum_{i} m_{i} \hat{m}_{i} $ by $ c_{j,\alpha} = \frac{1}{\sqrt{N}} \sum_{\mathbf{k}} c_{\mathbf{k},\alpha} e^{i \mathbf{k} \cdot \mathbf{R}_j} $, we obtain~\cite{Nayak2000}
	\begin{equation}
		\sum_{i} m_{i} \hat{m}_{i} = \sum_{\mathbf{k}, \mathbf{k}', \alpha} \sqrt{3}  m_{\mathbf{k}'-\mathbf{k}}  c_{\mathbf{k}',\alpha}^{\dagger} \lambda_8 c_{\mathbf{k},\alpha},
		\label{eq:magnetic_term}
	\end{equation}
	where $m_{\mathbf{k}'-\mathbf{k}}$ is the Fourier component and $\lambda_8 = \frac{1}{\sqrt{3}} \operatorname{diag}(1,1,-2)$ is the eighth Gell-Mann matrix~\cite{Georgi2000}. Similarly, the CDW term $\sum_{i} d_{i} \hat{d}_{i}$ is given by
	\begin{equation}
		\sum_{i} d_{i} \hat{d}_{i} = \sum_{\mathbf{k}, \mathbf{k}', \alpha} 3  d_{\mathbf{k}'-\mathbf{k}}  c_{\mathbf{k}',\alpha}^{\dagger} \lambda_0 c_{\mathbf{k},\alpha},
		\label{eq:cdw_term}
	\end{equation}
	where $d_{\mathbf{k}'-\mathbf{k}}$ is the Fourier component and $\lambda_0= \operatorname{diag}(1,1,1)$.
	
	Substituting Eqs.~\eqref{eq:magnetic_term}\eqref{eq:cdw_term} into the effective Hamiltonian~\eqref{eq:eff_hamiltonian} and taking
	the logarithm of Eq.~\eqref{eq:eff_SE}, we write the effective action in the Matsubara frequencies as
	\begin{equation}
		\begin{aligned}
			S_{E}[m, d] = & \int_{x} \left[ \frac{m_{i}^2}{2(\gamma-s+\frac{1}{3})} + \frac{ d_{i}^2}{2(\gamma+s+\frac{2}{3})} \right] \\
			& - \mathrm{Tr} \ln \big[ (-i \omega_n + \epsilon_{\mathbf{k}}) \delta_{\mathbf{k},\mathbf{k}'} \\
			& - (\sqrt{3}m_{\mathbf{k}'-\mathbf{k}} \lambda_8 + 3d_{\mathbf{k}'-\mathbf{k}} \lambda_0) \big],
		\end{aligned}
		\label{eq:matsubara_action}
	\end{equation}
	with $\int_{x} = \sum_{i} \int_{0}^{\beta} d\tau$, $\epsilon_{\mathbf{k}} = - 2\tilde{t} \sqrt{\cos^2 k_x + \cos^2 k_y}$, and $\tilde{t}=t/|U|$. For both orders with a wave vector $\mathbf{Q}=(\pi,\pi)$, we have~\cite{Nayak2000}
	\begin{align}
		m_{\mathbf{k}'-\mathbf{k}} & = \frac{1}{2} m (\cos k_x + \cos k_y) \delta_{\mathbf{k}', \mathbf{k} + \mathbf{Q}}, \label{eq:m_form_factor} \\
		d_{\mathbf{k}'-\mathbf{k}} & = \frac{1}{2} d (\cos k_x + \cos k_y) \delta_{\mathbf{k}', \mathbf{k} + \mathbf{Q}}. \label{eq:d_form_factor}
	\end{align}
	
	The GL free energy expansion is derived through introducing the non-interacting Green's function $ G_0(\mathbf{k}) = (i\omega_n - \epsilon_{\mathbf{k}})^{-1} $ and the mean-field operator $ V_{\mathbf{k},\mathbf{k}'} = -(\sqrt{3}m_{\mathbf{k}'-\mathbf{k}}\lambda_8 + 3d_{\mathbf{k}'-\mathbf{k}} \lambda_0) $. By expanding the effective action to fourth order, we obtain 
	\begin{equation}
		f = F - F_0 = \frac{a_m}{2} m^2 + \frac{u_m}{4} m^4 + \frac{a_d}{2} d^2 + \frac{u_d}{4} d^4 + \frac{\zeta}{2} m^2 d^2,
		\label{eq:gl_free_energy}
	\end{equation}
	where the GL coefficients are given by 
	\begin{equation}
	\begin{aligned}
		a_m & = \frac{1}{\gamma-s+1/3} + 3 \Pi(\mathbf{Q}),  \\
		a_d & = \frac{1}{\gamma+s+2/3} + 9 \Pi(\mathbf{Q}),  \\
		u_m & = 81 \Gamma(\mathbf{Q}, \mathbf{Q}),  \\
		u_d & = 6561 \Gamma(\mathbf{Q}, \mathbf{Q}),  \\
		\zeta & = 729 \Gamma(\mathbf{Q}, \mathbf{Q}), \label{eq:zeta_coeff}
	\end{aligned}
	\end{equation}
	with $\Pi(\mathbf{Q}) = \mathrm{Tr}[ \mathbf{\Delta}_{\mathbf{k}+\mathbf{Q}} G_0(\mathbf{k} + \mathbf{Q}) \mathbf{\Delta}_{\mathbf{k}} G_0(\mathbf{k}) ]$, $\Gamma(\mathbf{Q}, \mathbf{Q}) = \mathrm{Tr}[ (\mathbf{\Delta}_{\mathbf{k}+\mathbf{Q}})^2 G_0^2(\mathbf{k} + \mathbf{Q}) (\mathbf{\Delta}_{\mathbf{k}})^2 G_0^2(\mathbf{k}) ]$, and $\mathbf{\Delta}_{\mathbf{k}} = \cos k_x + \cos k_y$.
	
	To determine the constraints on GL coefficients required for order coexistence ($m \neq 0$, $d \neq 0$), we solve the saddle-point equations $\partial f / \partial m = 0$ and $\partial f / \partial d = 0$, and get the coupled equations
	\begin{equation}
		d^2 = \frac{-a_d - \zeta m^2}{u_d}, \quad m^2 = \frac{-a_m - \zeta d^2}{u_m}. 
		\label{eq:GL_coexist}
	\end{equation}
	The intrinsic symmetry constraint that the N\'eel order and the CDW order shall exhibit the same spatial modulation naturally leads to
	\begin{equation}
		\zeta^2 = u_m u_d.
		\label{eq:symmetry_constraint}
	\end{equation}
	Substituting Eq.~\eqref{eq:symmetry_constraint} into Eq.~\eqref{eq:GL_coexist}, we obtain the reduced constraint
	\begin{equation}
		a^2_m u_d = u_m a^2_d. 
		\label{eq:reduced_condition}
	\end{equation}
	Solving Eq.~\eqref{eq:reduced_condition}, we derive the parameter constraints
	\begin{equation}
		9\frac{1}{\gamma-s+1/3}+3\frac{1}{\gamma+s+2/3}=-2\times3\times9\Pi(\mathbf{Q}) ,
	\end{equation}
	or\begin{equation}
		2\gamma+4s+5/3=0. \label{constraint}
	\end{equation}
	with $s=|U^{\prime}|/3|U| \in (0,1/3]$. To obtain physical solutions for the conditions of order coexistence, the equations Eq.~\eqref{eq:symmetry_constraint}-~\eqref{constraint} must be satisfied. Under these constraints, Eq.~\eqref{eq:GL_coexist} takes the form
	\begin{equation}
		\frac{m^2}{\sqrt{u_d}} + \frac{d^2}{\sqrt{u_m}} = -\frac{a_d}{u_d\sqrt{u_m}} = -\frac{a_m}{u_m\sqrt{u_d}}.
		\label{eq:order_relation}
	\end{equation}
	The existence of solutions to Eq.~\eqref{eq:order_relation} necessitates $-a_m > 0$, $-a_d > 0$, $u_m > 0$, and $u_d > 0$.
		\begin{figure}[t]
		\centering
		\includegraphics[width=0.98\linewidth]{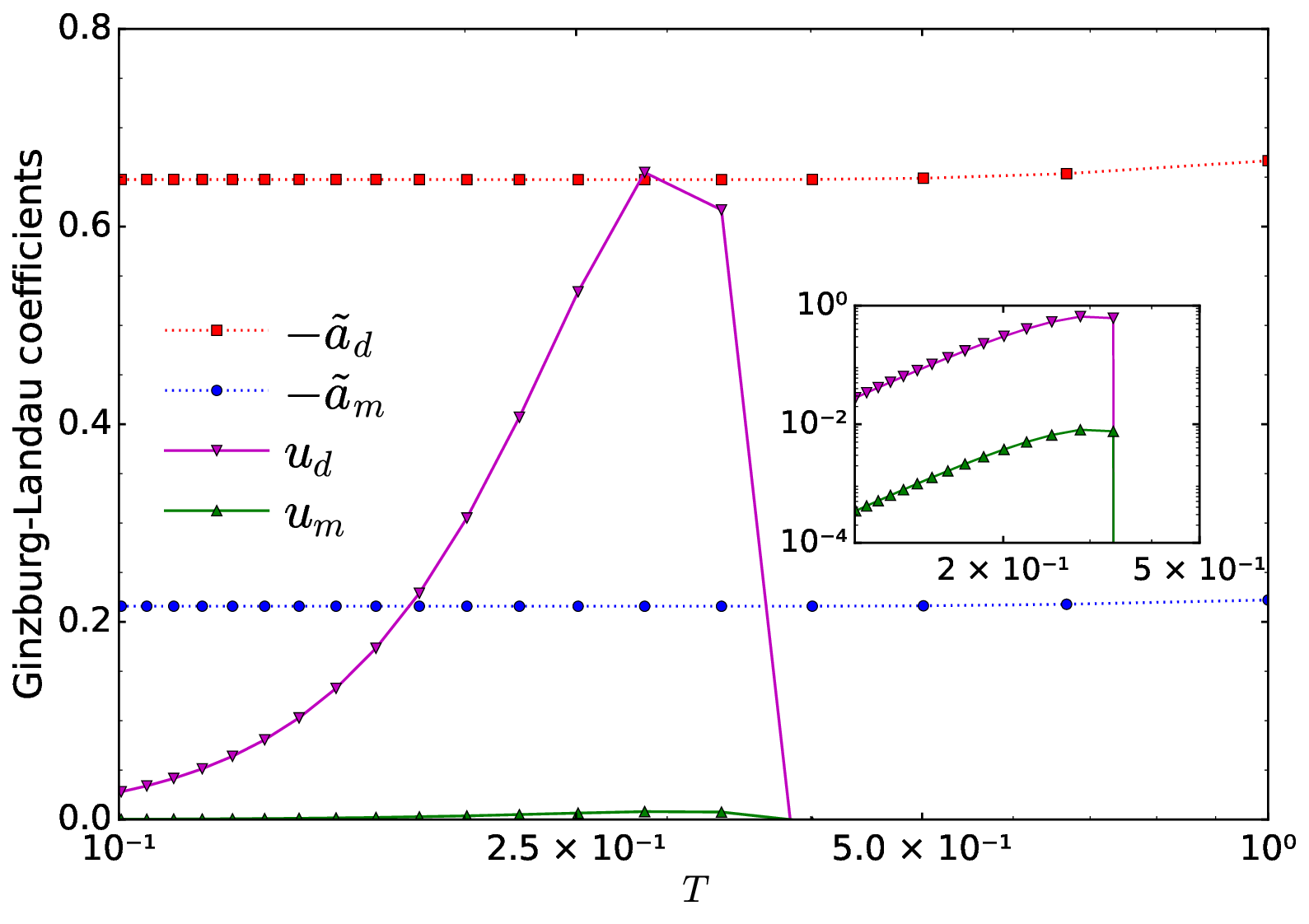}
		\caption{
			The temperature dependence of GL coefficients at $s=1/6$, $\gamma=-7/6$ (corresponding to $|U|=6$, $|U^{\prime}|=3$).  Rescaled coefficients $\tilde{a}_d$, $\tilde{a}_m$ and unmodified coefficients $u_d$, $u_m$ are shown. Sign changes in $u_m$ and $u_d$ signal simultaneous vanishing of coexisting CDW and N\'eel orders, consistent with DQMC results. The cutoff of Matsubara frequency $\omega_n$ with $|n| \leq 20$ and lattice size $L=80$ are used.
		}
		\label{fig:GL}
	\end{figure}
	
	We now plot the temperature dependence of GL coefficients (defined in Eq.~\eqref{eq:zeta_coeff}) in Fig.~\ref{fig:GL} at $s=1/6$, $\gamma=-7/6$ (corresponding to DQMC parameters $|U|=6$, $|U^{\prime}|=3$ and satisfying the constraint given by Eq.~\eqref{constraint}). For visual comparison, GL coefficients $a_d$ and $a_m$ are rescaled using $\tilde{a}_\alpha = \frac{a_\alpha}{1.5\max(|a_d|)}, \alpha \in \{d,m\}$, while $u_d$ and $u_m$ retain their unmodified values. This rescaling preserves the sign of GL coefficients but modifies their magnitudes for visualization clarity. From Fig.~\ref{fig:GL}, we conclude that the conditions of order coexistence ($-a_m > 0$, $-a_d > 0$, $u_m > 0$, $u_d > 0$) are satisfied throughout the low-temperature regime. The sign changes of coefficients $u_m$ and $u_d$ with increasing temperature represent simultaneous vanishing of coexisting CDW and N\'eel orders, consistent with DQMC results shown in Fig.~\ref{orders}. Crucially, our GL analysis reveals that the stable coexistence requires that the N\'eel ordering is originated from the CDW background and the SU(3) symmetry is broken. The qualitative agreement between the GL analysis and DQMC simulations confirms the dominant role of trionic states and a special mechanism in the coexisting CDW and N\'eel orders.
	\section{Conclusion and Discussion}
	\label{sec:conclusion}
	
	In summary, we have performed the sign-problem-free DQMC simulations to investigate the trion ordering in the half-filled attractive three-color Hubbard model on a $\pi$-flux square lattice. The color-dependent Hubbard attractions induce coexisting CDW and N\'eel orderings that survive up to the same melting temperature. As expected, N\'eel order on the $\pi$-flux square lattice is much stronger than that on the honeycomb lattice, since more hopping channels enhance charge fluctuations and then give rise to more off-site trions. Even in the presence of a $\pi$ flux, off-site trions are not long-range correlated, which means the off-site trion has no preferential spatial orientation. Within the framework of phenomenological GL theory, we demonstrate that color-dependent Hubbard interactions are the condition for inducing coexisting CDW and N\'eel orders.

	Coexistence of charge and magnetic orders have long been a major research focus in condensed matter physics. In the context of ultracold fermions, our work suggests a different mechanism for coexisting CDW and N\'eel orders, where the N\'eel order arises from the fluctuations of the CDW order and manifests as spatial modulation of off-site trions without any preferential orientation. In addition, our QMC results are asymptotic accurate, which provide a benchmark for future studies of the attractive three-color Hubbard model.

Finally we propose a scenario for observing the coexisting CDW and N\'eel orders through fluorescence imaging in cold atom experiment. In this approach, the charge and spin/color distributions can be simultaneously probed in optical lattices by performing separate imaging of each spin state \cite{Koepsell2020imag,Yan2022imag,Mongkolkiattichai2025imag}. Specifically in the three-color system, the in-plane atomic motion is frozen by a rapid increase of the lattice depth (achieved by increasing laser intensity), and then the color-$1$ state is removed to facilitate Stern-Gerlach separation \cite{Mongkolkiattichai2025imag}. Subsequently, the lattice depth is further increased, and meanwhile a vertical magnetic field gradient is applied to spatially split color-$2$ state and color-$3$ state into two sheets for simultaneous fluorescence imaging \cite{Koepsell2020imag,Yan2022imag,Mongkolkiattichai2025imag}. If the CDW order and the N\'eel order coexist, fluorescence images should simultaneously display (i) a similar staggered pattern on both sheets (i.e., bright spots tend to appear on site $i$ with an even value of $i_{x}+i_{y}$), characterizing the CDW order; and (ii) stronger staggered modulation of color-2 state (i.e. $M_{2}>M_{3}$), characterizing the N\'eel order. 
	\acknowledgments
This work is financially supported by the National Natural Science Foundation of China under Grants No. 12574298 and No. 11874292. We acknowledge the support of the Supercomputing Center of Wuhan University.
	
	\bibliographystyle{custom-apsrev}
%

\end{document}